\documentclass[twocolumn,twoside]{IEEEtran}
\usepackage{graphicx}
\usepackage{times}

\begin{document}
\title{ICN enabling CoAP Extensions for IP based IoT devices}

\author{\IEEEauthorblockN{Nikos Fotiou\IEEEauthorrefmark{1}, George Xylomenos\IEEEauthorrefmark{1}, George C. Polyzos\IEEEauthorrefmark{1},\\
Hasan Islam\IEEEauthorrefmark{2}, Dmitrij Lagutin\IEEEauthorrefmark{2}, Teemu Hakala\IEEEauthorrefmark{3}, Eero Hakala\IEEEauthorrefmark{3}}

\IEEEauthorblockA{\IEEEauthorrefmark{1}Athens University of Economics and Business, Greece. Email: \{fotiou,xgeorge,polyzos\}@aueb.gr}

\IEEEauthorblockA{\IEEEauthorrefmark{2}Aalto University, Finland. Email: firstname.lastname@aalto.fi}

\IEEEauthorblockA{\IEEEauthorrefmark{3}Ell-i open source co-operative, Finland. Email: temmi@iki.fi, eero.hakala@ell-i.org}}

\maketitle

\begin{abstract}
The Constrained Application Protocol (CoAP) and its extensions, such as observe and group communication, offer the potential for developing novel IoT applications. However, a full-fledged CoAP-based application requires delay-tolerant communication and support for multicast: since these properties cannot be easily provided by existing IP networks, developers cannot take full advantage of CoAP, preferring to use HTTP instead. In this demo we show how proxying CoAP traffic over an ICN network can unleash the full potential of CoAP, simultaneously shifting overhead and complexity from the (constrained) endpoints to the network.  
\end{abstract}

\begin{IEEEkeywords}
CoAP, Experimentation, Namespaces
\end{IEEEkeywords}

\section{Introduction}
The Constrained Application Protocol (CoAP)~\cite{rfc7252} has been called the ``HTTP for the
Internet of Things'' (IoT), as it allows CoAP clients to retrieve or set resources from CoAP
servers implemented in constrained devices (the Things). In contrast to HTTP however, it is
implemented over UDP and it allows for delayed responses, e.g., a CoAP client may request the
value of a resource that is not yet ready; it will initially receive an acknowledgment for its request 
and when the resource becomes available it will receive the appropriate response. Various CoAP
extensions enable novel applications, departing even further from the traditional
one request-one response model. For instance, the CoAP observe~\cite{rfc7641} extension allows CoAP clients
to \emph{observe} resources and receive a notification everytime their state changes (i.e.,
very similar to the publish-subscribe communication paradigm).
CoAP group communication is another CoAP extension defined in RFC 7390~\cite{rfc7390}. This extension
allows clients to retrieve or set resources from a \emph{group} of servers e.g., retrieve the temperature 
from all sensors of a building, turn on  all the lights of a factory, etc. Group names follow the structure
of legacy CoAP URIs, nevertheless they can be overloaded with application specific semantics
(e.g., a CoAP GET request for the group  \emph{coap://floor3.building6/temperature} may result in all temperature
sensors in floor 3 of building 6 to send a record). 

Despite its potential, CoAP is mostly used in its basic form 
(one request-one immediate response). The main reason for this is that a full fledged CoAP deployment 
is cumbersome, mostly due to the IP underlay. For example, in order to support delay-tolerant messaging or
publish-subscribe communication (i.e., CoAP observe) CoAP servers should maintain extensive state.
Similarly, RFC 7390 suggests that CoAP group communication
could be implemented by using IP multicast, with DNS mapping group names (included in the 
CoAP URI requests) into the appropriate IP multicast address: with this approach
CoAP servers should implement IP multicast and, for each group, a specific IP multicast address 
should be assigned (and configured in CoAP endpoints).

In this demo, we show how the POINT Information-Centric Networking (ICN) architecture can be leveraged, so 
that IP endpoints that implement only core CoAP, can benefit from CoAP and its extensions.
 
\begin{figure*}
\includegraphics[width=0.80\linewidth]{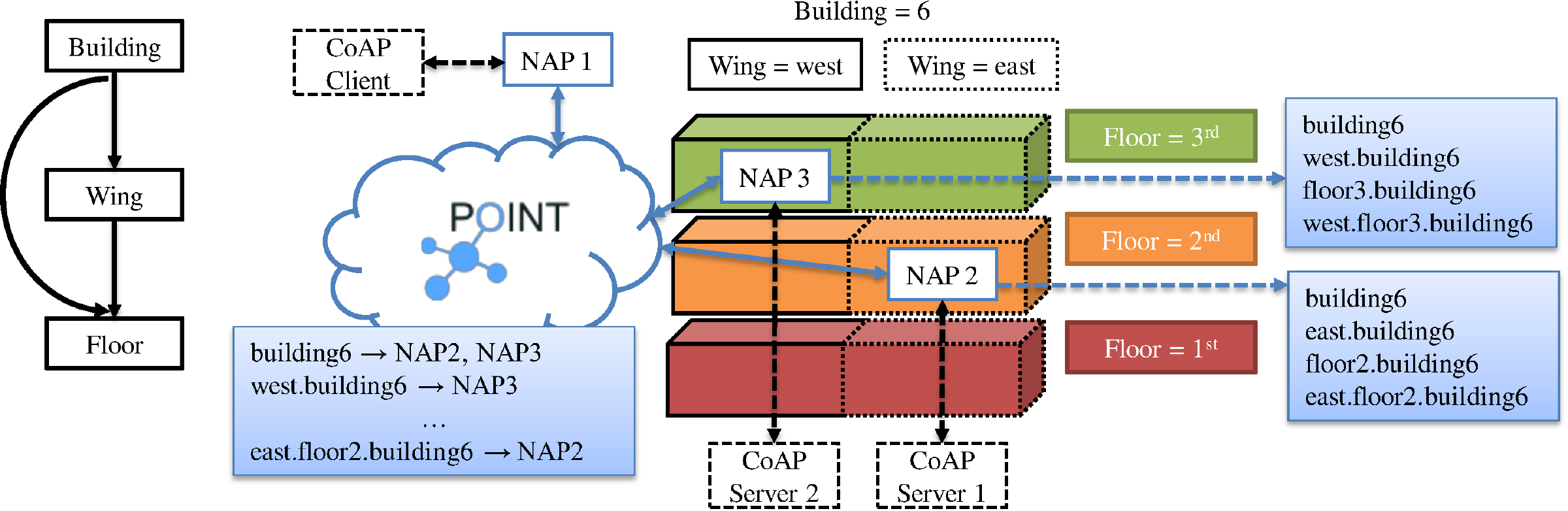}
\caption {High level overview of a POINT-enabled building management system.}
\label{fig:example}
\end{figure*}
\subsection{The POINT architecture}
The POINT architecture  allows standard IP traffic to be run over an ICN core network; the ICN core is typically  deployed  at a single network provider~\cite{Tro2015}. POINT's core ICN network is implemented
 using the Publish-Subscribe Internet (PSI) ICN architecture~\cite{Xyl2012},  a publish-subscribe architecture,
where users interested in receiving specific content subscribe to it, while content owners 
advertise their content and, if requested, publish it (i.e., they transfer it to the subscribers, by default through multicast).
The POINT architecture provides a number of handlers for existing 
IP-based protocols (e.g., HTTP, CoAP, and basic IP) that map the underlying protocols onto appropriate named objects
within the ICN core. Therefore, existing applications can benefit from ICN's features by forwarding their (legacy) 
traffic through Network Attachment Points (NAPs), where these protocols handlers  are implemented.

Figure~\ref{fig:example} gives a high level overview of a POINT network with CoAP handlers enabled in its NAPs.
NAPs connected to CoAP servers learn the URIs of the available resources and send subscription messages to  the ICN core
indicating their interest in receiving CoAP requests (encapsulated into POINT objects).
Similarly, whenever a NAP connected to CoAP clients (possibly, via the Internet) receives a CoAP request, it encapsulates it into a POINT content
item and advertises it in the ICN network. This advertisement will result in the content item being forwarded
to an appropriate NAP;  this NAP then will decapsulate the CoAP request and will forward it to the appropriate CoAP server. 
CoAP enabled NAPs can aggregate requests for the same CoAP resource, shifting this way state management from CoAP servers
to NAPs.
\section{Enabling CoAP extensions over POINT}
\subsection{Approach}
The POINT approach for supporting the CoAP observe extension is detailed in~\cite{isl2017}.
In a nutshell, the POINT NAPs aggregate requests for the same resource, hence from the CoAP server's
perspective only a single CoAP client is visible. Moreover, update notifications are transmitted using
multicast, thus conserving network resources. 
POINT leverages its information-centrism and its inherent support for multicast to support seamless 
and hassle-free group communication among CoAP endpoints.
In particular, it takes advantage of PSI's name structure in order to organise group ``attributes'';
then it assigns values to these attributes to construct group names, and maps these names into the appropriate PSI \emph{scopes}.

In order to illustrate this concept we consider the case of a building management system (depicted in Figure~\ref{fig:example}).
In this case there are CoAP servers located inside buildings and each server is attached to a NAP. Buildings are numbered with a \emph{building} number, and then subdivided in
\emph{wings} and \emph{floors}; these are the possible group attributes, which are hierarchically organised
as shown in the left part of Figure~\ref{fig:example}.
A CoAP client may send a request to a group of CoAP servers; the
group name is created by assigning ``values'' to (some of) the specified attributes, e.g., by setting $building=building6$,  
$wing=west$, and $floor=floor3$ the group name \emph{floor3.west.building6} is constructed.
POINT NAPs are configured with values for the specified attributes,
so that by using these values they can
construct all possible group names e.g., a NAP located in building6, west, $3^{rd}$ floor, 
creates the names \emph{building6}, \emph{building6.west},
\emph{floor3.building6}, and \emph{floor3.west.building6}.
Then, each NAP subscribes to the ICN content identifier that corresponds to each name.

Using this scheme, a CoAP request for a group (for example, \emph{coap://floor2.building6/temperature)},
is encapsulated into a POINT content item and is advertised in the ICN network
using as an identifier the FQDN of the CoAP server as specified in the request's URL (i.e.,
in our example `floor2.building6'); all NAPs that have subscribed to this identifier will receive that
item, will decapsulate the CoAP request and will forward it to the corresponding CoAP servers. Following a similar 
approach, CoAP responses are encapsulated into a POINT content item and are forwarded to the
appropriate NAPs and eventually to interested CoAP clients. 
\subsection{Key contributions}
The benefits of POINT to CoAP in general, are highlighted in~\cite{coaphandler}.
Similarly, the benefits of POINT to CoAP observe are highlighted in~\cite{isl2017}.
When it comes to CoAP group communication,
our approach enables issuing requests to groups of CoAP servers that implement the standard version of the CoAP protocol
(i.e., they do not support RFC 7390).
As a result, CoAP servers do not have to implement IP multicast.
Moreover, there is no need for modifications to DNS.
CoAP servers are oblivious to group names, since names are handled by the NAPs,
thus Things management becomes much easier. 
The ICN core makes group name administration easier:
new attributes can be easily added to the namespace without affecting already deployed NAPs.
Moreover, group names do not have to be mapped a priori to a lower layer network address.
\section*{Acknowledgments}
This research was supported by the EU funded H2020 ICT project POINT, under contract 643990.


\end{document}